\documentclass[twocolumn,astrosymb]{aastex701}

\usepackage{amsmath}
\usepackage{multirow}

\newcommand{\Liso}{L_{\gamma,\mathrm{iso}}}
\newcommand{\Eiso}{E_{\gamma,\mathrm{iso}}}
\newcommand{\Eisok}{E_{K,\mathrm{iso}}}
\newcommand{\flux}{\mathrm{erg}\,\mathrm{cm}^{-2}\,\mathrm{s}^{-1}}
\newcommand{\photonflux}{\mathrm{ph}\,\mathrm{cm}^{-2}\,\mathrm{s}^{-1}}
\newcommand{\fluence}{\mathrm{erg}\,\mathrm{cm}^{-2}}

\newcommand{\vastafterglows}{8\pm5}

\newcommand{\swiftargusrate}{16\pm1}

\newcommand{\fermiargusrate}{51\pm5}
\newcommand{\fermiarguseff}{(26\pm3)\%}
\newcommand{\starburstargusrate}{84\pm6}

\newcommand{\moonbeamargusrate}{90\pm9}

\newcommand{\fermidsarate}{86\pm8}
\newcommand{\fermidsaeff}{(44\pm3)\%}
\newcommand{\starburstdsarate}{158\pm11}

\newcommand{\moonbeamdsarate}{160\pm14}

\newcommand{\fracPrepeak}{18\%}
\newcommand{\fracFifteen}{71\%}          
\newcommand{\fracBaseNotDay}{23\%}   
\newcommand{\fracDayNotBase}{26\%}   
\newcommand{\fracBaseAndDay}{30\%}   

\newcommand{\lsstFactorOneDet}{1.4}       
\newcommand{\lsstFactorTwoDetSame}{2.3}   
\newcommand{\lsstFactorFade}{5.5}         

\newcommand{\unc}{
    Department of Physics and Astronomy, 
    University of North Carolina at Chapel Hill, 
    Chapel Hill, NC 27599-3255, USA
}
\newcommand{\umn}{
    School of Physics and Astronomy, 
    University of Minnesota, 
    Minneapolis, Minnesota 55455, USA
}
\newcommand{\usyd}{
    Sydney Institute for Astronomy, 
    School of Physics, 
    University of Sydney, 
    Sydney, NSW 2006, Australia
}
\newcommand{\ozgrav}{
    ARC Centre of Excellence for Gravitational Wave Discovery (OzGrav), 
    Hawthorn, Victoria, 3122, Australia
}
\newcommand{\mcwilliams}{
    McWilliams Center for Cosmology and Astrophysics,
    Department of Physics,
    Carnegie Mellon University,
    5000 Forbes Avenue, Pittsburgh, PA 15213, USA
}
\newcommand{\lsu}{
    Department of Physics \& Astronomy, 
    Louisiana State University, 
    Baton Rouge, LA 70803, USA
}
\newcommand{\harvard}{
    Center for Astrophysics, 
    Harvard \& Smithsonian, 
    60 Garden Street, Cambridge, MA 02138-1516, USA
}

\newcommand{\usra}{
    Science and Technology Institute, 
    Universities Space Research Association, 
    Huntsville, AL 35805, USA
}
\begin{document}

\title{Prospects for GRB Afterglow Discovery with the Eric and Wendy Schmidt Observatory System}

\author[orcid=0009-0006-7990-0547,gname='James' ,sname='Freeburn']{James Freeburn}
\affiliation{\usyd}
\affiliation{\ozgrav}
\affiliation{\unc}
\email[show]{james.freeburn@sydney.edu.au}

\author[0000-0003-0699-7019]{Dougal Dobie} 
\affiliation{\usyd}
\affiliation{\ozgrav}
\email{d.dobie@sydney.edu.au}

\author[0000-0002-8935-9882]{Akash Anumarlapudi} 
\affiliation{\unc}
\email{akasha@unc.edu}

\author[0000-0002-8977-1498]{Igor Andreoni}
\affiliation{\unc}
\email{igor.andreoni@unc.edu}

\author[0000-0002-9700-0036]{Brendan O'Connor}
\altaffiliation{McWilliams Fellow}
\affiliation{\mcwilliams}
\email{boconno2@andrew.cmu.edu}


\author[0000-0002-2942-3379]{Eric Burns}
\affiliation{\lsu}
\email{ericburns@lsu.edu}

\author[0000-0001-8544-584X]{Jonathan Carney}
\affiliation{\unc}
\email{jcarney@unc.edu}

\author[0000-0002-8935-9882]{Hank Corbett} 
\affiliation{\unc}
\email{hcorbett@ad.unc.edu}

\author[0000-0002-8262-2924]{Michael W. Coughlin}
\affiliation{\umn}
\email{cough052@umn.edu}

\author[0000-0002-0587-7042]{Adam Goldstein}
\affiliation{\usra}
\email{agoldstein@usra.edu}

\author[0000-0001-9229-8833]{Anna Tartaglia}
\affiliation{\harvard}
\email{atartaglia@fas.harvard.edu}

\author[0000-0002-5814-4061]{V. Ashley Villar}
\affiliation{\harvard}
\affiliation{The NSF AI Institute for Artificial Intelligence and Fundamental Interactions, USA}
\email{ashleyvillar@cfa.harvard.edu}

\begin{abstract}

Two time domain surveys, recently funded as part of the Eric and Wendy Schmidt Observatory System; the Argus Array, in the optical, and the Deep Synoptic Array (DSA), in the radio, will transform gamma-ray burst (GRB) science via the serendipitous discovery of hundreds of GRB afterglows per year.  In this work, we simulate DSA and Argus observations of GRB afterglows.  We find that, of the long-duration GRBs (LGRBs) detected by the Fermi Gamma-ray Burst Monitor, $\fermiarguseff$ will yield afterglow detections with Argus and $\fermidsaeff$ with DSA, corresponding to rate of $\fermiargusrate$ and $\fermidsarate$ per year respectively.  We also compute rates for both upcoming and proposed GRB monitors; the forthcoming StarBurst Multi-messenger Pioneer, with $\starburstargusrate$ detections per year in Argus and $\starburstdsarate$ detections per year in DSA and the Moon Burst Energetics All-sky Monitor (MoonBEAM) concept, with $\moonbeamargusrate$ per year in Argus and $\moonbeamdsarate$ per year in DSA.  The observatory system will detect also 118$\pm$10 optical and 199$\pm$17 radio afterglows per year, independent of GRB triggers, exceeding the current annual rate with global follow-up.  Afterglow counterparts to short-duration GRBs, originating from neutron star mergers, will be detected at 5--10\% of the LGRB afterglow rate, which is promising for multi-messenger detections of gravitational wave sources and constraining the neutron star merger rate.  The Argus Array, with its second--minute cadence, will detect afterglows before they peak $\sim\fracPrepeak$~of the time which will dramatically increase the sample of observed reverse shock and prompt optical emission.

\end{abstract}


\section{Introduction}

Gamma-ray bursts (GRBs) originate from highly relativistic jets thought to be launched by either the collapse of massive stars \citep{Woosley1993,1998Natur.395..670G,2003Natur.423..847H} or the mergers of compact objects \citep{Eichler89,Gehrels05,at2017gfo5}. GRBs are the most powerful explosions in the Universe and are probes of extreme physics \citep[e.g.][]{2013PhRvD..87l2001V,2017ApJ...848L..13A, 2018PhRvL.121p1101A} and cosmology \citep[e,g.][]{2017Natur.551...80K, 2008ApJ...683L...5Y}.

Previously, in optical and radio wavelengths specifically, targeted follow-up has been relied on for detecting afterglow emission.  Significant progress has been made with dedicated follow-up programs with Swift's Ultraviolet/Optical Telescope \citep{2005SSRv..120...95R,2009ApJ...690..163R,2009MNRAS.395..490O}, Gamma-Ray burst Optical and Near-infrared Detector \citep[GROND;][]{2008PASP..120..405G,2012A&A...548A.101N}, the RAPid Telescopes for Optical Response \citep[RAPTOR;][]{2002SPIE.4845..126V}, the Robotic Optical Transient Search Experiment \citep[ROTSE;][]{2009ApJ...702..489R} and the Jansky Very Large Array and the Australia Telescope Compact Array \citep{2003AJ....125.2299F,2012ApJ...746..156C}, among others \citep{2009ApJ...702..489R,2009AJ....137.4100K,2012ExA....33..173K}.  Alternatively, multiple afterglows have been discovered serendipitously in synoptic surveys with the Palomar Transient Factory \citep{2013ApJ...769..130C}, the Zwicky Transient Facility \citep{2021ApJ...918...63A,2022ApJ...938...85H}, the Rapid ASKAP Continuum Survey \citep{2020PASA...37...48M,2021MNRAS.503.1847L} and the Transiting Exoplanet Survey Satellite \citep{2024ApJ...963...89R,2024ApJ...972..162J}.  However, limitations on sensitivity, sky coverage or cadence have limited the quantity and density of early-time sampling for these events.

Schmidt Sciences has recently funded the Eric and Wendy Schmidt Observatory System, which will consist of four observatories; the Argus Array \citep{2022PASP..134c5003L}, the Deep Synoptic Array \citep[DSA;][]{2019BAAS...51g.255H}, Large Fiber Array Spectroscopic Telescope \citep[LFAST;][]{2022SPIE12182E..1UA} and Lazuli Space Observatory \citep{2026arXiv260102556R}. DSA and Argus are discovery facilities that will provide the capability to serendipitously discover afterglows, surpassing triggered approaches in the quantity of afterglows they detect and characterize while utilizing substantially fewer resources. LFAST and Lazuli will provide the rapid follow-up that is necessary for understanding these events -- LFAST will provide optical spectroscopy while Lazuli will provide optical and infra-red photometry and spectroscopy. Combined, these four facilities provide a comprehensive end-to-end system capable of discovering and rapidly characterizing GRB afterglows.

Optical afterglow emission is typically dominated by the forward shock formed as the GRB jet interacts with the circumburst medium.  However, bright, fast-fading optical flashes have been detected, preceding the forward shock emission and their mechanism is the subject of debate \citep[e.g.][]{1999ApJ...517L.109S,2009ApJ...691..723B,2011A&A...528A..15G,2013ApJ...774..114J,2014Sci...343...38V,2017Natur.547..425T,2019ApJ...879L..26F,2023NatAs...7..843O,1999MNRAS.306L..39M,2005Natur.435..178V,2008Natur.455..183R,2010ApJ...719L..10B,2026arXiv260305608J}.  With its second--minute timescale cadence, sensitivity and 8,000 square degree field-of-view, the Argus Array will routinely, serendipitously observe these optical flashes and aid in resolving its dominant mechanism.  Additionally, afterglows' fast fade rates make spectroscopy costly, often requiring interruptive observations with large aperture telescopes \citep{2019A&A...623A..92S}.  Argus will enable rapid spectroscopic follow-up more frequently by localizing the burst within minute-timescales, while also providing a critical measure of its optical brightness.  This early spectroscopy is fruitful as it yields fine structure absorption lines which can be used to directly probe the environment surrounding the GRB \citep[e.g.][]{2004A&A...419..927V,2005ApJ...634L..25C,2006ApJ...648...95P,2007A&A...468...83V} and unambiguously measure redshifts, constraining the energetics and progenitors of hostless GRBs \citep[][]{2002AJ....123.1111B,2010ApJ...722.1946B,2013ApJ...769...56F,2014MNRAS.437.1495T,2022MNRAS.515.4890O,2025MNRAS.537.2061F}.  With the absorption features imparted by the interstellar medium (ISM), the intergalactic medium (IGM) and intervening absorbers at high redshifts, the evolution of cosmic star formation \citep[e.g.][]{1997ApJ...486L..71T,2000ApJ...536....1L,2004ApJ...609..935Y,2008ApJ...683L...5Y, 2009ApJ...705L.104K, 2009ApJ...691..182S}, chemical enrichment \citep[e.g.][]{2004A&A...419..927V, 2005A&A...442L..21S,2006A&A...451L..47F,2006Natur.440..184K,2007ApJS..168..231P,2010A&A...523A..36D} and the reionization of the IGM can be measured \citep[e.g.][]{1998ApJ...501...15M,2000ApJ...536....1L,2006PASJ...58..485T}.

The DSA will have unprecedented survey speed, surpassing existing facilities by two orders of magnitude, and enabled by its extreme continuum sensitivity (600\,nJy in 1 hour across 0.7-2\,GHz) and relatively large field of view (2.5\,$\deg$ full-width half-maximum at 1.35\,GHz). It will carry out an all-sky time domain survey, probing 3 times more sky volume at a substantially higher cadence than the best radio time domain surveys conducted to date \citep{2019BAAS...51g.255H,2026arXiv260222739D}. The observed population of radio afterglows typically peak tens of days post-burst \citep[e.g.][]{2012ApJ...746..156C}, but there is likely a strong bias towards early-rising bursts due to the limited amount of observing time available -- most searches cease monitoring after several weeks without a detection. However, the standard GRB models predict afterglows that may not peak for hundreds, or even thousands, of days post-burst depending on the properties of the circumburst medium and the microphysics parameters of the jet. There is likely an unobserved population of bursts that are not detectable with the current resource-constrained observing programs but will be key to understanding the overall GRB population. The DSA all-sky survey solves this by providing coverage of every GRB in the survey footprint across the survey lifetime, regardless of the detection date, ultimately providing the first sample of radio afterglows that is unbiased by observability and resource constraints. The DSA will not eliminate the need for dedicated follow-up (e.g. it cannot provide early-time coverage, nor the frequency coverage necessary to determine spectral evolution), but the sensitivity will surpass most of the targeted L and S-band searches conducted to-date, which can reach a noise level of $\sim 7\,\mu$Jy, but typically are $>10\,\mu$Jy \citep[e.g.][]{2024MNRAS.533.4435R,2025ApJ...982...42S}.

In this paper we simulate a population of GRBs and Argus and DSA observations of their corresponding afterglows. We show that both facilities will expand, by orders of magnitude, the rate of discovering well-sampled, optical and radio afterglows as part of their planned surveys.  Here, we are focused on the afterglows accompanying both long (LGRBs) and short-duration (SGRBs) GRBs and leave simulations of GRB-associated supernovae and kilonovae and `orphan' afterglows \citep[e.g.][]{2020MNRAS.491.5852A,2022ApJ...938...85H,2024MNRAS.531.4836F} to future work. 

In \S\ref{sec:telescopes}, we describe the technical specifications of Argus, DSA and the current and planned GRB detectors simulated in this work.  We then describe our methods for simulating afterglow observations with these facilities in \S\ref{sec:method}, including generating a synthetic population of GRBs and their afterglows in \S\ref{sec:population} and survey simulations for Argus and DSA in \S\ref{sec:argus_sims} and \S\ref{sec:sims_DSA} respectively.  Measurements of afterglow detection rates and efficiencies from a range of GRB detectors and comparisons with other time-domain surveys are presented in \S\ref{sec:results_discussion}.  We then summarize our findings in \S\ref{sec:conclusion}.

Throughout this work, we assume a flat $\Lambda$CDM cosmology where $H_0=70$\,km\,s$^{-1}$\,Mpc$^{-1}$, $\Omega_m=0.3$ and $\Omega_\Lambda=0.7$.

\section{Telescope Description and Capabilities}\label{sec:telescopes}

\subsection{Argus Array}

The Argus Array is an upcoming optical synoptic survey telescope with a modular design.  It comprises of an array of 1,212 0.28\,m telescopes mounted with 102-MPix CMOS detectors.  This yields an 8\,m effective diameter, 1 arcsecond resolution and an instantaneous field-of-view of 8,000 square degrees.  CMOS detectors enable a very high observational cadence owing to their ms-timescale read-out times\footnote{Argus' specifications and data products are described at \url{https://argus.unc.edu}.}.

While the survey strategy is subject to change, we assume the following strategy throughout this work.  The northern sky, $\delta>-20^\circ$, is imaged at a baseline cadence of 1\,minute in dark and grey time and 1\,second in bright time reaching as deep as $g\sim$20.5 and 17.1\,AB mag respectively.  The baseline survey will utilize alternating g- and r-band filters, permanently fixed to each telescope with a 2:1 ratio.  Argus will track the sky in 15\,minute intervals, followed by a shift in pointing.  A typical visible patch of sky for a given night, will therefore receive a continuous cycle of $g$-band observations for 30\,minutes followed by 15\,minutes of $r$-band observations.

A difference imaging and transient detection pipeline, using on-site computing resources, is run on the data in real-time.  Transient alerts will from this pipeline will be released with a latency of nine seconds in second cadence mode and one minute in minute cadence mode.  Among the data products will be alerts at the baseline cadence in addition to deeper, coadded data of 15\,minutes ($g\sim22.1$), 1\,hour ($g\sim22.9$), 1\,night ($g\sim23.9$) and 1\,week ($g\sim24.7$).  For a single night, Argus' footprint will span 21,000 square degrees and in a year, it will cover 29,650 square degrees \citep{2022SPIE12189E..10C}.

Argus is a promising facility for afterglow detection owing to its very high cadence compared to other optical synoptic surveys.  The Zwicky Transient Facility \citep[ZTF;][]{2019PASP..131a8002B} and the Vera C. Rubin Observatory's Legacy Survey of Space and Time \citep[LSST;][]{2019ApJ...873..111I} image the visible sky at a cadence of a day or longer.  Argus will image the entire visible sky every night with much of its footprint receiving continuous monitoring at minute or second cadence for much of the night.  Additionally, Argus' depth eclipses other wide-field surveys with high cadence such as the Transiting Exoplanet Survey Satellite \citep[TESS;][]{2015JATIS...1a4003R} and Evryscope \citep{2015PASP..127..234L}.

\subsection{Deep Synoptic Array}

The Deep Synoptic Array (DSA) will consist of $1650\times 6.15$\,m with instantaneous frequency coverage spanning 0.7--2.0\,GHz, a maximum baseline of 20\,km. The resulting astrometric precision should be $\lesssim$0.3", sufficient for clear host association and any multi-wavelength follow-up.  The bulk of its operations will be dedicated to carrying out an all-sky time domain survey, nominally covering the sky North of $-31\deg$ (corresponding to $\sim 31,000\,\deg^2$ to a sensitivity of $2\,\mu$Jy (between $0.7-2$\,GHz) on a $\sim 4$ month cadence. 

These specifications drastically surpass the current most sensitive surveys -- the extragalactic component of the ASKAP Variables And Slow Transients Survey covers $\sim$12,300$\,\deg^2$ to a sensitivity of $\sim$250$\,\mu$Jy at 888\,MHz on a cadence of 2 months \citep[][]{2026arXiv260222739D}, the Caltech-NRAO Stripe Survey covers $270\,\deg^2$ to a sensitivity of $\sim 40\,\mu$Jy at 3\,GHz \citep{2016ApJ...818..105M}, while the VLA Sky Survey covers $\sim$33885$\,\deg^2$ to a sensitivity of $140\,\mu$Jy at 3\,GHz on a cadence of 32 months \citep{2020PASP..132c5001L,2025usnc.conf..226M}. Searches for extragalactic synchrotron transients, including GRB afterglows, with these surveys have thus far been hindered by a combination of sensitivity, observing cadence and total sky coverage \citep[e.g.][]{2023MNRAS.523.4029L,2025MNRAS.538.2676G}. The DSA all-sky survey will reach a sensitivity two orders of magnitude deeper than VLASS while covering a similar total area of sky, while the cadence will be comparable to that of VAST, allowing for monitoring of the evolution of transients throughout time without requiring dedicated follow-up observations.

\subsection{GRB detectors}

We compute rates for Swift/BAT and Fermi/GBM in addition to future facilities; StarBurst and MoonBEAM \citep{2023arXiv230816293F,2024Univ...10..187B}. Their specifications are listed in Table~\ref{tab:instrument_specs}.

\begin{deluxetable*}{lcccccr}
\tablewidth{0pt}
\tablecaption{Existing and future GRB detector sensitivities and specifications used in this work for computing rates.\label{tab:instrument_specs}}
\tablehead{
\colhead{Instrument} &
\colhead{Energy band (keV)} &
\colhead{LGRB sensitivity threshold} &
\colhead{SGRB sensitivity threshold} &
\colhead{$\Omega_\gamma$ (sr)} &
\colhead{$D_\gamma$}
}
\startdata
Swift/BAT & 15--150 & $S>2.8\times10^{-8}\,\flux$ & $S>1.1\times10^{-7}\,\flux$ & 2.2 & 78\% \\
Fermi/GBM & 30--500 & $S>5.7\times10^{-8}\,\flux$ &  $S>1.6\times10^{-7}\,\flux$ & $3\pi$ & 85\% \\
StarBurst & 30--500 & $S_{\mathrm{ph}}>0.25\,\photonflux$ & $S_{\mathrm{ph}}>0.58\,\photonflux$ & 8 & 85\% \\
MoonBEAM & 30--500 & $S_{\mathrm{ph}}>0.73\,\photonflux$ & $S_{\mathrm{ph}}>1.5\,\photonflux$ & $4\pi$ & 98\% \\
\enddata
\end{deluxetable*}

\subsection{Swift/BAT}\label{sec:swift}

The Burst Alert Telescope on-board the Neil Gehrels Swift Observatory \citep[Swift/BAT;][]{2005SSRv..120..143B} is a hard X-ray telescope with a coded mask, enabling arcminute-scale localizations.  It provides near real-time triggering and alerts for multi-wavelength follow-up.  With Swift/BAT, we assume an LGRB rate of 71.65\,yr$^{-1}$ and an SGRB rate of 7.59\,yr$^{-1}$ for 15--150\,keV occurring in the 10\% coded region comprising 2.2\,sr with a duty cycle of 78\% following the results of \citet{2016ApJ...829....7L}.  We compute an effective energy flux thresholds by setting it at a value which recovers observed GRB detection rates in our simulations described in \S\ref{sec:grb_pop}.  This yields thresholds of $2.8\times10^{-8}\,\flux$ for LGRBs and $1.1\times10^{-7}\,\flux$ for SGRBs.  These values are consistent with observations reported in \citet{2016ApJ...829....7L}.

\subsection{Fermi/GBM}

The Fermi Gamma-ray Burst Monitor \citep[Fermi/GBM][]{2009ApJ...702..791M} is a wide-field high energy telescope, sensitive in the 1-1000\,keV energy band.  It observes the entire unocculted sky, which we assume corresponds to a field-of-view of $3\pi$\,sr with a duty cycle of 85\%, as measured by \citet{2020ApJ...893...46V}.  for Fermi/GBM \citep{2020ApJ...893...46V}.   We follow the same procedure as in \S\ref{sec:swift} to compute the sensitivity for this work with an LGRB rate of 195.8\,yr$^{-1}$ and an SGRB rate of 39.5\,yr$^{-1}$ \citep{2020ApJ...893...46V}.  Here, we compute the flux sensitivity threshold in the 50--300\,keV trigger band.  This yields $5.7\times10^{-8}\,\flux$ for LGRBs and $1.6\times10^{-7}\,\flux$ for SGRBs, consistent with the results of \citet{2020ApJ...893...46V}.

\subsubsection{StarBurst}\label{sec:starburst}

The StarBurst Multi-messenger Pioneer \citep{2024HEAD...2140602K} is a wide-field, highly sensitive GRB monitor, planned for launch in Spring of 2028 in order to overlap with the fifth observing run of the International Gravitational Wave Network.  It has a design similar to Fermi/GBM accessing the entire unocculted sky in LEO but achieves a 400--500\% effective area compared to Fermi/GBM due to its detector design \citep{2024HEAD...2140602K,2024Univ...10..187B}. 

Sensitivities are generally reported as photon flux thresholds integrated at 64\,ms and 1024\,ms in the 50--300\,keV band.  For LGRBs, the integrated sensitivity at 1024\,ms is used as LGRBs have durations that exceed this trigger timescale.  For SGRBs, however, shorter duration bursts will be triggered in the 64\,ms data while bursts appproaching the canonical $T_{90} = 2$\,s boundary will be more likely to be triggered in the 1024\,ms window.  We therefore approximate the estimate the SGRB detection threshold by taking the midpoint between these two thresholds.  We assume the reported GRB sensitivity in \citet{2024HEAD...2140602K} of $S_{\mathrm{ph}}>0.25\,\photonflux$ at 1024\,ms and $S_{\mathrm{ph}}>0.9\,\photonflux$ at 64\,ms for the 30--500\,keV energy band.

\subsubsection{MoonBEAM}

The proposed Moon Burst Energetics All-sky Monitor \citep[MoonBEAM;][]{2023arXiv230816293F} mission is designed to provide continuous observing of the entire sky, with an order of magnitude more sensitivity than previous instruments. This would be accomplished by an orbit far outside of Low Earth Orbit (LEO). Such a design achieves a field-of-view comprising $>99$\% of the sky and a duty cycle of $>98\%$. It will localize GRBs in a manner similar to Fermi/GBM with degree-scale localization regions \citep{2023arXiv230816293F,2024Univ...10..187B}.  

We utilize simulated efficiency curves in the 30--500\,keV energy band for MoonBEAM from \citet{2023arXiv230816293F} and at compute the detection threshold at the 50\% efficiency level.  This yields $S_{\mathrm{ph}}>2.3\,\photonflux$ at 64\,ms and $S_{\mathrm{ph}}>0.73\,\photonflux$ at 1024\,ms.  We adopt the same convention as \S\ref{sec:starburst}; assuming the 1024\,ms threshold for LGRBs and the midpoint between the 64\,ms and 1024\,ms thresholds for SGRBs.

\section{Survey Simulation}\label{sec:method}

By simulating observations of afterglows, we aim to estimate the detection rates of the multi-wavelength counterparts to GRBs with Argus and DSA.  We therefore require a synthetic population of short and long-duration GRBs to determine their detectability with current and future GRB monitors, their corresponding afterglow emission and a realistic set of simulated observations in each survey.  With the simulated observations of our synthetic population, we can then compute detection efficiencies and rates.

\subsection{Population synthesis}\label{sec:population}

\subsubsection{LGRBs}\label{sec:grb_pop}

We synthesize 10$^6$ GRBs with redshifts up to $z=10$ and isotropic-equivalent gamma-ray luminosities ($\Liso$) from the GRB formation rate, $\rho(z)$, and luminosity function $\Phi(z,\Liso)$ derived in \citet[][GS22]{2022ApJ...932...10G}, respectively.  These are given by;
\begin{equation}   
    \rho(z) = \rho_0 \frac{(1+z)^{p_{z,1}}}{1 + \left(\dfrac{1+z}{p_{z,2}}\right)^{p_{z,3}}}
\end{equation}
where $\rho_0=79$\,Gpc$^{-3}$\,yr$^{-1}$, $p_{z,1}=3.33$, $p_{z,2}=3.42$ and $p_{z,3}=6.21$,
\begin{equation}
    \begin{split}    
    \Phi(\Liso, z) \propto
    \begin{cases}
        \left(\dfrac{\Liso}{L_{\gamma,\mathrm{iso,b}}}\right)^{-a_1} & \Liso \le L_{\gamma,\mathrm{iso,b}} \\
        \left(\dfrac{\Liso}{L_{\gamma,\mathrm{iso,b}}}\right)^{-a_2} & \Liso > L_{\gamma,\mathrm{iso,b}}
    \end{cases}
    \end{split}
\end{equation}

\begin{equation}
    L_{\mathrm{\gamma,iso,b}} = L_{\gamma,\mathrm{iso,0}} (1+z)^{\delta}
\end{equation}

where $\Phi(\Liso, z)$ is normalized with the inverse integral of the luminosity function at a given redshift, $a_1=0.97$, $a_2=2.21$, $L_{\gamma,\mathrm{iso,0}}=10^{52.02}$\,erg and $\delta=0.64$.  Following GS22, we assume a Band function \citep{1993ApJ...413..281B} spectrum for each GRB described by,
\begin{equation}
N(E) \propto
\begin{cases}
E^{\alpha}\, \exp\!\left(-\dfrac{E}{E_0}\right),
& E < E_b \\[6pt]

E^{\beta},
& E \ge E_b
\end{cases}
\end{equation}

\begin{equation}
    E_0 = \frac{E_p}{\alpha + 2},
\end{equation}
where $E_p$ is the energy at peak of $E \times N(E)$. The parameters $E_p$, $\alpha$, $\beta$ in addition to the isotropic-equivalent energy release of the GRB, $\Eiso$, the half-jet opening angle for the prompt $\gamma$-ray emission, $\theta_{\gamma,j}$, and the initial Lorentz factor of the jet $\Gamma_0$ are drawn from the correlation functions derived in GS22.

For a given energy band, we then compute observer frame peak flux ($S$), peak photon flux ($S_{\mathrm{ph}}$) and fluence ($F$) with 
\begin{equation}
    S = \frac{1}{I}\frac{\Liso}{4\pi d_L^2}\int_{E_{\mathrm{min}}}^{E_{\mathrm{max}}}E\, N_{\mathrm{obs}}(E)\, dE
\end{equation}
\begin{equation}
    S_{\mathrm{ph}} = \frac{1}{I}\frac{\Liso}{(1+z)4\pi d_L^2}\int_{E_{\mathrm{min}}}^{E_{\mathrm{max}}}N_{\mathrm{obs}}(E)\, dE
\end{equation}
\begin{equation}
    F=\frac{1}{I}\frac{(1+z)\,E_{\rm iso}}{4\pi d_L^2}\int_{E_{\mathrm{min}}}^{E_{\mathrm{max}}}E\, N_{\mathrm{obs}}(E)\, dE
\end{equation}
where $N_{\rm obs}(E)$, is the GRB spectrum shifted to the observer frame, $E_{\rm \min}$ and $E_{\rm \max}$ are the bounds of the energy band, $d_L$ is luminosity distance and $I$ is the integral of the GRB spectrum in the rest frame from from 1--$10^4$\,keV.

We draw viewing angles, $\theta_v$, from a probability density proportional to $\sin\theta_v$ and assess an on-axis viewing angle to be those that satisfy $\sin\theta_v \leq \max(\sin(\theta_{\gamma,j}), 1/\Gamma_0)$.  For this work, we only consider the 48,712 on-axis bursts, excluding off-axis bursts from our original sample of 10$^6$.

\subsubsection{SGRBs}

We generate 10,000 on-axis SGRBs using the population synthesis model described in \citet[G16;][]{2016A&A...594A..84G}.  Namely, we use a GRB formation rate of the form \citep{2001MNRAS.326..255C},
\begin{equation}
    \Psi(z) = \frac{1 + p_1 z}{1 + (z / z_p)^{p_2}}
\end{equation}
with $p_1=2.8$, $p_2=3.5$ and $z_p=2.3$ and draw $E_p$ values from the following probability density function,
\begin{equation}
    \phi(E_p) \propto
\begin{cases}
\left( \dfrac{E_p}{E_{p,b}} \right)^{-a_1} & E_p \leq E_{p,b} \\
\left( \dfrac{E_p}{E_{p,b}} \right)^{-a_2} & E_p > E_{p,b}
\end{cases}
\end{equation}
where $\alpha_1=0.53$, $\alpha_2=4$ and $E_{p,b}=1600$\,keV.  We then draw $\Eiso$ and $\Liso$ values from their resultant correlations with $E_p$ in G16 and assume a consistent band function spectrum with indices $\alpha=-0.6$ and $\beta=-2.5$.  Flux, fluence and photon flux values are computed with the method described in \S\ref{sec:grb_pop}.

\subsubsection{LGRB afterglows}\label{sec:afterglow_pop}

To generate a realistic population of afterglows, as counterparts to our synthetic population of LGRBs, we require a distribution of microphysical parameters that can reproduce the observational properties of detected radio and optical afterglows.  We first compile a set of observed afterglow light curves in X-ray, optical and radio wavelengths.  Using this sample, we select a subset of light curves and generate synthetic GRBs that match the observed redshift distribution. Finally, we fit afterglow microphysical parameters to the observed light curves, producing posterior distributions consistent with afterglow observations.  We note that, for the purposes of this work, we are not aiming to generate a `true' distribution of afterglow parameters.  There are observational biases here that cannot be accounted for.  We are simply aiming to generate a set of afterglow light curves which recreate the variety, evolution and brightness of observed afterglow light curves in order to estimate detection efficiencies and rates. Below we describe our process in detail.

We utilize observed LGRB afterglow light curves in $R_c$-band, corrected for Galactic extinction, compiled in \citet{2006ApJ...641..993K,2010ApJ...720.1513K,2011ApJ...734...96K,2012A&A...548A.101N}, observed by the Jansky Very Large Array at 8.46\,GHz compiled in \citet{2026arXiv260220522G} \citep[from][]{2012ApJ...746..156C} and at 1\,keV observed by the Swift X-ray telescope \citep[Swift/XRT;][]{2007A&A...469..379E} which have been corrected for absorption.  

The Optically Unbiased GRB Host Survey \citep[TOUGH;][]{2012ApJ...756..187H} catalog, is a complete sample which includes measured redshifts for 53 GRBs.  The TOUGH catalog has three selection criteria which are applied to GRB properties; $T_{90}>2$\,s, an on-board Swift/BAT trigger and a detection of and X-ray afterglow within 12\,hr of the GRB.  GS22 show that a cut of $F>10^{-6}\fluence$ from $15-150$\,keV on their population synthesis model yields a redshift distribution consistent with that of TOUGH.  As a result, for fitting LGRB afterglows, we select synthetic LGRBs with $F>10^{-6}\fluence$ from $15-150$\,keV from and remove observed optical and radio afterglows that do not have a counterpart detected by Swift/XRT, following the selection criteria for TOUGH.  

We favor light curves with early detections to mitigate a bias towards brighter events.  We therefore apply an additional requirement that the first detection occurs within $2\times10^{-3}$\,d in X-rays, $10^{-2}$\,d in the optical following the GRB.  Due to the comparatively long duration and variability of radio rise times, we do not apply this cut to the radio light curves. For fitting LGRB afterglows, we obtain a sample of 310, 72 and 27 light curves in X-rays, optical and radio respectively.

To ensure our synthetic population is consistent with the flux density evolution of observed afterglows, we interpolate five time steps in each light curve and take the log median and log standard deviation of the flux densities at each time-step.  The time-ranges are chosen to maximize temporal coverage while ensuring at least 25 percent of the light curves have coverage at each time-step.  These time steps are geometrically sampled $10^{-2}$--5\,d post-burst in the optical, $2\times10^{-3}$--1\,d post-burst in X-rays and 5--10$^2$\,d post-burst in radio.

To this observational data, we can fit synthetic distributions of light curves.  For this task, we use \texttt{VegasAfterglow} \citep{2026JHEAp..5000490W} owing to its flexibility and high performance treatment of synchrotron self-absorption, jet Lorentz factor and reverse shocks.  We draw 100 GRBs satisfying $F>10^{-6}\,\fluence$ for 15--150\,keV from our synthetic GRB population described in \S\ref{sec:grb_pop} because GS22 show that this cut yields a redshift distribution consistent with the TOUGH catalog.   We take the log-mean and log-standard deviations of the simulated flux densities at the same time-steps as with the observational data.

Given our optical dataset is in a single band-pass, we opt to utilize the complete sample of host-galaxy extinction measurements gathered in \citet{2013MNRAS.432.1231C}. We fit a log-normal probability density function yielding a best-fit function centered at $\ln A_v=-0.97$ with a variance of $\sigma_{\ln A_v}=0.86$.  For each GRB in our population, we draw an extinction value from this probability density and assume the extinction law derived in \citet{2007ApJ...663..320F} for both host galaxy and Galactic extinction.

We assume a gamma-ray efficiency ($\eta_{\gamma}$) of 15\%, following \citet{2013MNRAS.433.2107N,2016MNRAS.461...51B,2017MNRAS.472.3161B}, which is given by
\begin{equation}
    \eta_{\gamma} = 0.15 = \frac{\Eiso}{\Eiso + \Eisok}
\end{equation}
where $\Eisok$ is the on-axis isotropic equivalent kinetic energy in the jet which is calculated from the $\Eiso$ values drawn in \S\ref{sec:grb_pop}.  Given the collapsar progenitors of LGRBs \citep[e.g.][]{1998Natur.395..670G,1999ApJ...524..262M,2003Natur.423..847H}, it is expected that they should occur in a wind medium \citep{1998ApJ...497L..17S,1999ApJ...520L..29C,2000ApJ...536..195C}.  However, the most LGRB afterglows are consistent with an ISM medium \citep{2011A&A...526A..23S}.  We therefore assume a uniform interstellar medium (ISM), a tophat jet and an on-axis viewing angle ($\theta_v=0$).  Furthermore, we decouple the afterglow jet-opening $\theta_c$ from the $\gamma$-ray jet-opening angles, $\theta_{\gamma,j}$, drawn in \S\ref{sec:grb_pop}.  This is to both account for possible differences in the beaming factor between the afterglow and GRB and to allow for a wider variety of light curve morphologies in our fit.  We fit for the ISM density, $-4 < \log_{10} n_0 < 2$; the energy fraction stored in the jets electrons, $-3.0 < \log_{10}\epsilon_e < -0.5$; the energy fraction stored in the magnetic field, $-5 < \log_{10}\epsilon_B< -1$; the afterglow jet-opening angle $-4 < \log_{10}\theta_c < -0.5$ and the electron power law index $2 < p < 3$.

\begin{figure*}
    \centering
    \includegraphics[width=1.3\columnwidth]{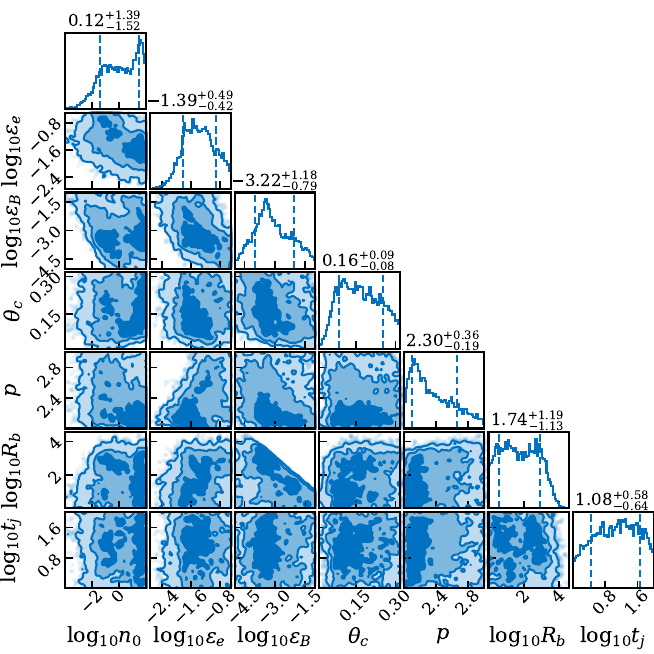}
    \caption{Posterior distribution of an MCMC fit of a synthetic LGRB afterglow population to a selection of observed afterglow flux density distributions at a range of frequencies and times.  The jet-opening angle, $\theta_c$, is in units of radians. Our synthetic population of afterglows are drawn from this posterior distribution.}
    \label{fig:afterglow_distribution_corner}
\end{figure*}

\begin{figure*}
    \centering
    \includegraphics[width=\linewidth,trim=0 6 0 0, clip]{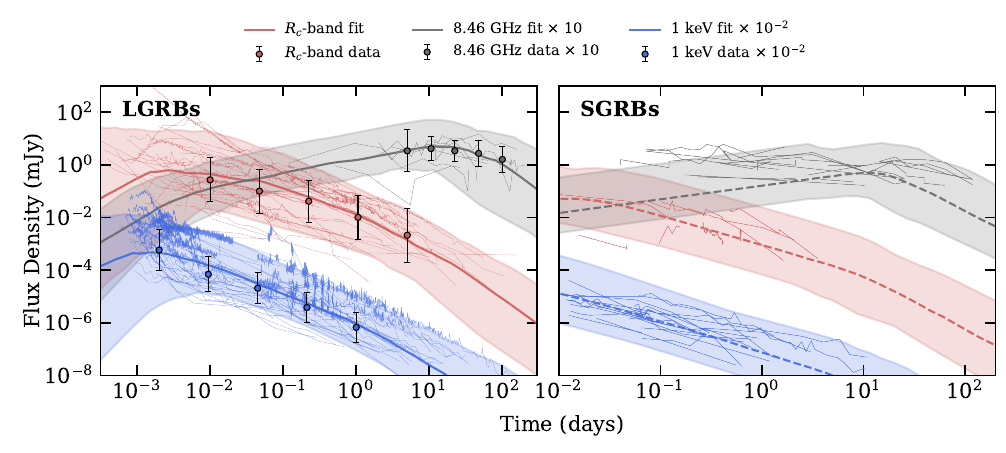}
    \caption{Observed and model light curves for both SGRBs and LGRBs.  Right panel: LGRB afterglow models resulting from the MCMC fit in Fig.~\ref{fig:afterglow_distribution_corner} and described in \S\ref{sec:afterglow_pop}.These are interpolated at specific time steps for 8.46\,GHz, $R_c$-band and 1\,keV and their distributions in log-space are shown as colored points, with error bars denoting their standard deviations.  Left panel: SGRB models are drawn from a manually selected distribution of parameters.  The thin lines denote individual observed light curves and  population models are shown with the thick colored lines denoting the log-mean flux density and the shaded regions denoting the 90\% percent confidence interval.}
    \label{fig:afterglow_distribution_fit}
\end{figure*}

To simulate reverse shock emission, we assume that $p$ and $\epsilon_e$ are the same for both forward and reverse shocks and fit for the magnetization parameter, $R_B = \epsilon_{\mathrm{rvs},B/\epsilon_B}$ and the duration of the central engine, $t_j$.  We uniformly sample $0>\log_{10}R_B>5$ and $0<\log_{10}t_j<2$ with a constraint of $\epsilon_{\mathrm{rvs},B}+\epsilon_{e}<1$. Reverse-shock emission is bright at early times, outside the temporal range of our observational data.  Following the results of \citet{2014ApJ...785...84J}, we therefore apply an additional constraint of $f_{\rm rvs>fwd}=(13\pm6)$\%, where $f_{\rm rvs>fwd}$ is the fraction of light curves with reverse shock emission that peaks brighter than the forward shock emission in $R_c$-band:
\begin{equation}
\ln \mathcal{L}_{\rm rvs}
= -\frac{1}{2}
\left[
\left(\frac{\Delta f_{\rm rvs>fwd}}{\sigma_{\rm rvs>fwd}}\right)^2
+ \ln \sigma_{\rm rvs>fwd}^2
\right]
\end{equation},
\begin{equation}
    \Delta f_{\rm rvs>fwd} = f_{\rm rvs>fwd,model} - f_{\rm rvs>fwd}.
\end{equation}
We minimize the Gaussian log-likelihood,
\begin{equation}
\ln \mathcal{L} = \ln \mathcal{L}_{\rm rvs} + \ln \mathcal{L}_{\rm fwd},
\end{equation}
\begin{equation}
    \mathcal{L}_{\rm fwd} = \frac{1}{2}\sum_i\left[
\frac{(\mu_{\mathrm{model},i} - \mu_{\mathrm{obs},i})^2}{\sigma_{\mathrm{tot},i}^2}
+ \ln\sigma_i^2
\right],
\end{equation}
\begin{equation}
\sigma_{\mathrm{tot},i}^2 = \sigma_{\mathrm{model},i}^2 + \sigma_{\mathrm{obs},i}^2,
\end{equation}
where $\mu_{i}$ and $\sigma_{i}$ are computed from a set of $N$ light curves,
\begin{equation}
\mu_{i} = \frac{1}{N} \sum_{j}^{N} \log_{10} f_j(\nu_i, t_i)
\end{equation}
\begin{equation}
    \sigma_{i}^2 = 
\frac{1}{N} \sum_{j}^{N} \left[ \log_{10} f_j(\nu_i, t_i) - \mu_i \right]^2
\end{equation}
with flux densities for light curve $j$, $f_j(\nu_i, t_i)$ at data point $i$ with a specific time, $t_i$, and frequency, $\nu_i$.  This was conducted using \texttt{bilby} \citep{2019ApJS..241...27A} with \texttt{emcee} \citep{2013PASP..125..306F} as a sampler.  The posterior distribution is shown in Fig.~\ref{fig:afterglow_distribution_corner} and the light curve distribution is shown in Fig.~\ref{fig:afterglow_distribution_fit}.

\subsubsection{SGRB afterglows}\label{sec:afterglow_pop_short}

Compared to LGRBs, SGRBs have a small number of observed afterglows due to their comparative faintness and rareity.  The method of fitting to observed flux distributions adopted in \S\ref{sec:afterglow_pop} is therefore ill advised.  We instead manually set parameters, informed by the light curves and estimates reported in \citet{2015ApJ...815..102F}.  Specifically, we assume $\theta_v=0$ \citep[see, e.g.,][]{OConnor2024} and $\eta_\gamma=0.15$ as in \S\ref{sec:afterglow_pop} but conversely, $A_v=0$, $\Gamma_0=500$, $\epsilon_e=0.1$ and $\epsilon_B=0.01$ for all SGRB events.  We uniformly draw values $-3 < \log n_0 < -1$, $\log_{10}(6^\circ) < \log \theta_c < \log_{10}(20^\circ)$ and $2.2 < p < 2.8$. These values span the range inferred for both circumburst density \citep{2015ApJ...815..102F,Oconnor2020} and jet half-opening angle \citep[][and references therein]{2015ApJ...815..102F,RoucoEscorial2023}. Reverse-shock emission is not simulated for SGRBs as there are too few examples in the literature to constrain its parameters.  In Fig.~\ref{fig:afterglow_distribution_fit}, we show the resultant distribution of afterglow light curves, simulated with \texttt{VegasAfterglow}, is consistent with a set of observed, Galactic extinction corrected, $R_c$-band light curves from \citet[][and references therein]{2015ApJ...815..102F}, 6--10\,GHz light curves from \citet[][and references therein]{2026arXiv260220522G} and absorption corrected 1\,keV light curves compiled from the Swift/XRT GRB catalog\footnote{\url{https://www.swift.ac.uk/xrt_curves/}} \citep{2007A&A...469..379E}.

\subsection{Argus Array}\label{sec:argus_sims}

We simulate the 5-year baseline Argus survey using \texttt{ArgusSim}, a simulation framework that models the full expected error budget of the instrument, including telescope optics, detector characteristics, and site-specific observing constraints. We assume a mid-latitude Northern Hemisphere site for the Argus Array site throughout, with typical weather patterns and seeing distributions modeled from continental sites in the desert southwest of North America, and nominal sensor performance (1.4 e$^-$ readout noise and 0.002 e$^-$ s$^{-1}$ pixel$^{-1}$ dark current). \texttt{ArgusSim} generates mock observing logs and noise budgets for a HEALPix \citep{2005ApJ...622..759G} grid of sky regions at a representative resolution. The sky background dominates the noise budget in the 1-minute cadence survey, and the simulation incorporates the effects of spatial and temporal variations in sky brightness due to scattered moonlight, zodiacal light, gegenschein, contributions from blended stars, artificial light, and airglow. The framework calculates an AB 5-$\sigma$ limiting magnitude at each epoch from atmospheric extinction, telescope and filter throughput, and detector quantum efficiency. 

\begin{figure}
    \centering
    \includegraphics[width=\linewidth,trim=10 0 20 10, clip]{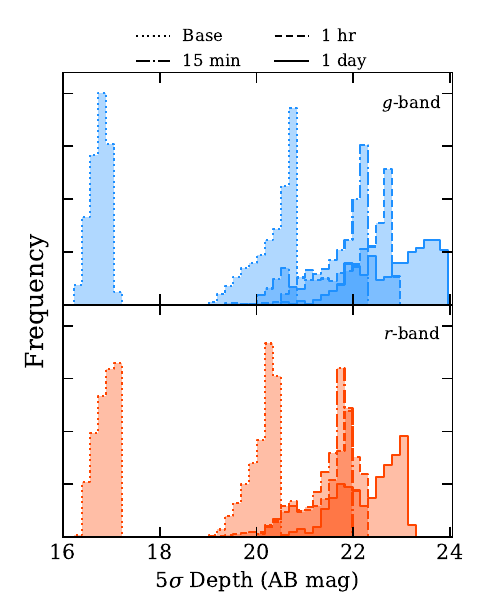}
    \caption{Distribution of simulated depths in Argus $g$ and $r$-band for a the baseline cadence in addition to observations stacked at 15 minute, one hour and day duration intervals.  The y-axis shows the from the density of observations at each depth bin and cadence, in arbitrary units.  The baseline cadence includes second cadence observations in bright time with median depths of $g=16.8$ and $r=16.9$\,AB mag and minute cadence data in dark and grey time with depths of $g=20.5$ and $r=20.2$\,AB mag.}
    \label{fig:depth_distribution}
\end{figure}

To maxmimize computational efficiency, we only use observations of a single, representative HEALPix region (6131).  This represents the limiting magnitudes and visits a single patch of sky will receive.  Any biases introduced by using a single HEALPix—-such as those related to observability, cadence or airmass—-are expected to be mitigated over the duration of the simulated five-year survey.  Additionally, Galactic extinction values are independently assigned by randomly sampling coordinates from the entire Argus footprint.  The resultant distribution of 5-$\sigma$ limiting magnitudes is shown in Fig.~\ref{fig:depth_distribution}.

We use \texttt{redback} \citep{2024MNRAS.531.1203S}, a software package for simulating and fitting electromagnetic transients, with the \texttt{SimulateOpticalTransient} class to simulate Argus observations of our population of afterglows modeled with \texttt{VegasAfterglow}.  We also only simulate events that satisfy $z<4.5$, simulating the Lyman break drop out at the Argus band passes.  This reduces the number of simulated events to 42,803.  Each event is assigned a random, uniformly distributed $t_0$ between the start and end of the simulated observations, a right ascension from 0--360$^\circ$ and a cosine-scaled declination from -20--90$^\circ$.  Galactic extinction is computed for each set of coordinates using the infrared dust map of the Milky Way \citep{1998ApJ...500..525S} calibrated in \citet{2011ApJ...737..103S}. Each transient is simulated for the baseline (second cadence in bright time and minute cadence in grey and dark time), 15 minute, one hour and day cadence.  For the baseline cadence, flux densities are modeled at the midpoint of the exposure and calculated by integrating spectrum, with host galaxy and Galactic extinction applied, across the transmission of each filter.  The flux densities for the coadded cadences are computed by taking the mean of the flux density measurements for each exposure in the coadd. The 5-$\sigma$ limiting magnitudes for the coadds are computed with $m_{\rm lim,coadd} = m_{\rm lim} + 2.5\log_{10}(\sqrt{N})$, where $m_{\rm lim}$ is the single frame depth and $m_{\rm lim,coadd}$ is the coadded depth.  A `detection' is considered when a simulated afterglow has a magnitude brighter than $m_{\rm lim}$ for the base cadence, and brighter than $m_{\rm lim,coadd}$ for the coadded observations.  For computing metrics, we consider all observations that have a magnitude error $>0.3$.

While an afterglow may be detected by Argus, for it to be readily identified in the alert stream, it would likely need to pass some basic filtering.  We therefore compute a set of metrics for each simulated Argus light curve which are described in Table~\ref{tab:metrics}.  These include metrics based on number of detections, fade-rate and multi-band coverage.  Fade rate is calculated by fitting a line for all detections from the light curve peak for each filter individually.

\begin{deluxetable}{lll}
\tablewidth{0pt}
\tablecaption{Metrics used to assess detectability in simulated Argus and DSA observations.\label{tab:metrics}}
\tablehead{
\colhead{Survey} &
\colhead{Metric name} &
\colhead{Description}
}
\startdata
\multirow{6}{*}{Argus} 
& \texttt{1det} & $\geq$ one detection \\
& \texttt{2det} & $\geq$ two detections in any filter \\
& \texttt{2det,band} & $\geq$ two detections in a single filter  \\
& \texttt{2det,multi-band} & $\geq$ two detections in multiple filters\\
& \texttt{fade} & $\geq$ a detectable fade rate of 0.3 magnitudes per day\\
& \texttt{fade,multi-band} & \texttt{fade} with $\geq$ one detection in multiple filters\\
\hline
\multirow{3}{*}{DSA} 
& \texttt{1det} & $\geq$ one detection \\
& \texttt{2det} & $\geq$ two detections \\
& \texttt{3det} & $\geq$ three detections \\
\enddata
\end{deluxetable}

\subsection{Deep Synoptic Array}\label{sec:sims_DSA}

A similar analysis was performed to estimate the detection fraction of GRB afterglows in the DSA all-sky survey. We simulated DSA observations using the nominal survey parameters \citep{2019BAAS...51g.255H} --- all-sky survey North of $-31\,\deg$ declination with a 4-month cadence and a sensitivity of $2\,\mu$Jy per 20\,min visit. Given the field of view ($\approx10.5$\,$\deg^2$), and the sky area coverage ($\sim$31,000$\deg^2$), we tiled the sky uniformly and chose the observation times to randomly span the cadence between multiple epochs.

We injected sources uniformly on the celestial sphere. For each source, we assign a $t_0$, chosen to be randomly distributed within one year after the survey start date. This allows us to sample both the early time behavior and the late time behavior ($>$\,yr) of the radio afterglows. We then sample the light curve at the time of the visits, given a sky position, and judge the detectability based on two simple criteria -- i) the position of the afterglow must be within the survey footprint, and ii) the flux density must be $>5\sigma$ at the time of the observation, where $\sigma$ is the survey sensitivity per observing epoch. For every source, we note the number of epochs in which it is detected and the peak flux density of the detection. We iterate this process 1,000 times to account for Poisson statistics and consider the mean detection rate across iterations. 

We note that scintillation is not simulated here and will likely affect how identifiable detected afterglows are and possibly their detectability \citep{2014PASA...31....8G}.  The afterglow population is fit to observations at 8.46\,GHz, higher than that of the DSA's frequency coverage.  Synchrotron self-absorption at 0.7--2\,GHz, therefore, may not be well constrained.   We also reiterate that an ISM environment is assumed.  While most radio afterglows are consistent with an ISM environment \citep{2011A&A...526A..23S} where the peak flux density at $\nu < \nu_m$ increases as t$^{1/2}$, wind environments have a constant peak flux with time \citep{2002ApJ...568..820G}.  This may result in an overestimation of afterglow rates with DSA, as they retain their brightness for longer in the ISM regime.

\subsection{Rates}

The rate of GRBs detectable in a given energy band and sensitivity, with afterglows detectable by either Argus or DSA per unit solid angle, is given by

\begin{equation}
\Re = N_{\rm detected} \times \frac{\Re_{\rm ref}}{N_{\rm ref}}
\end{equation}

where $N_{\rm detected}$ is the number of GRBs in our synthetic population that would have an afterglow detected in the given survey. The quantities $\Re_{\rm ref}$ and $N_{\rm ref}$ denote the reference event rate per unit solid angle and the number of events in the synthetic population satisfying the corresponding selection criteria used for the rate calibration. For LGRBs, $\Re_{\rm ref}=4.6\times10^{-3}$\,deg$^{-2}$\,yr$^{-1}$ and $N_{\rm ref}=569$, which is obtained from the BAT6 sample with $S_{\rm ph}>2.6\,\photonflux$ in the 15--150\,keV energy band \citep{2022ApJ...932...10G}.  For short GRBs, we use the Fermi/GBM SGRB detection rate of $\Re_{\rm ref}=9.7\times10^{-4}$\,deg$^{-2}$\,yr$^{-1}$ with $S_{\rm ph}>5$\,$\photonflux$ in the 10--1000\,keV energy band \citep{2016A&A...594A..84G}.

The all-sky detection rate, regardless of whether a GRB is detected, is given by
\begin{equation}
\Re_{\rm detection} = \Re \times \Omega
\end{equation}
and, imposing a GRB detection,
\begin{equation}
\Re_{\rm detection,GRB} = \Re \times \Omega_\gamma \times D_\gamma \times \Omega/\Omega_{\rm sky}
\end{equation}
where $\Omega_\gamma$ is the field-of-view of the given GRB detector, $D_\gamma$ is the duty cycle of the detector, and $\Omega/\Omega_{\rm sky}$ is the fraction of sky covered by the survey.

\section{Results \& Discussion}\label{sec:results_discussion}

\subsection{Rates}

\begin{figure}
    \centering
    \includegraphics[width=\linewidth]{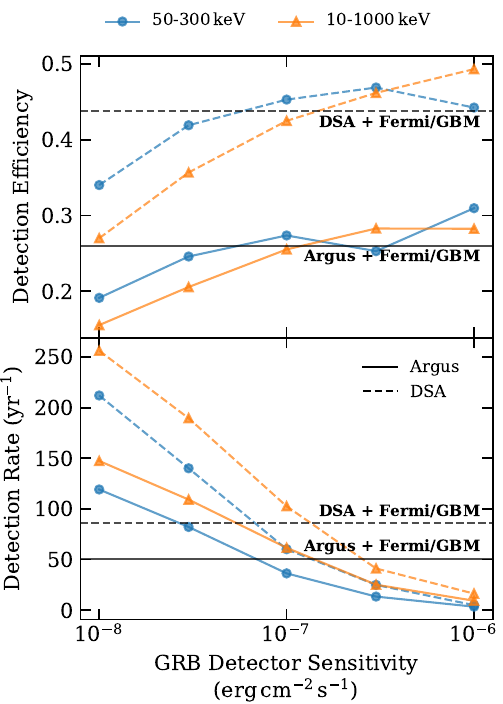}
    \caption{Detection efficiency and rates of afterglow detection with DSA and Argus from GRBs detected with a variety GRB energy bands and sensitivities.  We assume $\Omega_\gamma=3\pi$ and $D_\gamma=0.95$.  Top panel: the afterglow detection efficiency with GRB detector sensitivity. Bottom panel: afterglow detection rates with GRB detector sensitivity.  The afterglow detection efficiencies from Argus and DSA for GRBs detected with Fermi/GBM are plotted as black, horizontal lines for comparison.}
    \label{fig:instrument_detection_rates}
\end{figure}

With the procedure described in \S\ref{sec:afterglow_pop} and \S\ref{sec:afterglow_pop_short}, we compute efficiencies and rates of detection for afterglow counterparts to GRBs detected by the instruments listed in Table~\ref{tab:instrument_specs}.  Tables~\ref{tab:metrics_results_argus} and \ref{tab:metrics_results_argus_sgrb} show these measurements for Argus and Tables~\ref{tab:metrics_results_DSA} and \ref{tab:metrics_results_DSA_sgrb} show them for DSA.  We show the expected efficiencies and rates of detection with a range of flux sensitivity thresholds and energy bands in Fig \ref{fig:instrument_detection_rates}.  

\begin{deluxetable*}{c r r r r r}

\tablewidth{0pt}
\tablecaption{Detection metrics for simulated LGRB afterglows in Argus.\label{tab:metrics_results_argus}}
\tablehead{
\colhead{Metric} &
\colhead{Efficiency} &
\colhead{$N_{\rm detected}$} &
\colhead{$\Re$ (deg$^{-2}$ yr$^{-1}$)\,$\times10^3$} &
\colhead{$\Re_{\rm detection}$ (yr$^{-1}$)} &
\colhead{$\Re_{\rm detection,GRB}$ (yr$^{-1}$)}
}
\startdata
Swift/BAT & -- & 423 & $12.60\pm0.61$ & -- & $71\pm3$\\
\hline
\texttt{1det} & $26.3\pm2.1\%$ & 166 & $4.94\pm0.38$ & $136.8\pm10.6$ & $18.7\pm1.5$ \\
\texttt{2det} & $25.9\pm2.1\%$ & 163 & $4.85\pm0.38$ & $134.4\pm10.5$ & $18.3\pm1.4$ \\
\texttt{2det,band} & $24.4\pm2.1\%$ & 154 & $4.59\pm0.37$ & $127.0\pm10.2$ & $17.3\pm1.4$ \\
\texttt{2det,multi-band} & $24.9\pm2.1\%$ & 157 & $4.68\pm0.37$ & $129.4\pm10.3$ & $17.7\pm1.4$ \\
\texttt{fade} & $22.7\pm2.0\%$ & 143 & $4.26\pm0.36$ & $117.9\pm9.9$ & $16.1\pm1.3$ \\
\texttt{fade,multi-band} & $21.9\pm2.0\%$ & 138 & $4.11\pm0.35$ & $113.8\pm9.7$ & $15.5\pm1.3$ \\
\hline
Fermi/GBM & -- & 250 & $7.45\pm0.47$ & -- & $196\pm12$\\
\hline
\texttt{1det} & $30.9\pm2.9\%$ & 115 & $3.42\pm0.32$ & $94.8\pm8.8$ & $60.4\pm5.6$ \\
\texttt{2det} & $30.1\pm2.9\%$ & 112 & $3.34\pm0.32$ & $92.3\pm8.7$ & $58.9\pm5.6$ \\
\texttt{2det,band} & $28.5\pm2.9\%$ & 106 & $3.16\pm0.31$ & $87.4\pm8.5$ & $55.7\pm5.4$ \\
\texttt{2det,multi-band} & $29.3\pm2.9\%$ & 109 & $3.25\pm0.31$ & $89.9\pm8.6$ & $57.3\pm5.5$ \\
\texttt{fade} & $26.0\pm2.8\%$ & 97 & $2.89\pm0.29$ & $80.0\pm8.1$ & $51.0\pm5.2$ \\
\texttt{fade,multi-band} & $25.2\pm2.7\%$ & 94 & $2.80\pm0.29$ & $77.5\pm8.0$ & $49.4\pm5.1$ \\
\hline
StarBurst & -- & 608 & $18.11\pm0.73$ & -- & $404\pm16$\\
\hline
\texttt{1det} & $24.5\pm1.7\%$ & 222 & $6.61\pm0.44$ & $183.0\pm12.3$ & $99.0\pm6.6$ \\
\texttt{2det} & $23.7\pm1.7\%$ & 215 & $6.40\pm0.44$ & $177.2\pm12.1$ & $95.9\pm6.5$ \\
\texttt{2det,band} & $22.4\pm1.7\%$ & 203 & $6.05\pm0.42$ & $167.4\pm11.7$ & $90.6\pm6.4$ \\
\texttt{2det,multi-band} & $23.0\pm1.7\%$ & 208 & $6.19\pm0.43$ & $171.5\pm11.9$ & $92.8\pm6.4$ \\
\texttt{fade} & $20.9\pm1.6\%$ & 189 & $5.63\pm0.41$ & $155.8\pm11.3$ & $84.3\pm6.1$ \\
\texttt{fade,multi-band} & $20.2\pm1.6\%$ & 183 & $5.45\pm0.40$ & $150.9\pm11.2$ & $81.6\pm6.0$ \\
\hline
MoonBEAM & -- & 289 & $8.61\pm0.51$ & -- & $348\pm20$\\
\hline
\texttt{1det} & $31.6\pm2.7\%$ & 136 & $4.05\pm0.35$ & $112.1\pm9.6$ & $109.9\pm9.4$ \\
\texttt{2det} & $30.9\pm2.7\%$ & 133 & $3.96\pm0.34$ & $109.6\pm9.5$ & $107.5\pm9.3$ \\
\texttt{2det,band} & $28.6\pm2.7\%$ & 123 & $3.66\pm0.33$ & $101.4\pm9.1$ & $99.4\pm9.0$ \\
\texttt{2det,multi-band} & $30.0\pm2.7\%$ & 129 & $3.84\pm0.34$ & $106.3\pm9.4$ & $104.2\pm9.2$ \\
\texttt{fade} & $26.0\pm2.6\%$ & 112 & $3.34\pm0.32$ & $92.3\pm8.7$ & $90.5\pm8.6$ \\
\texttt{fade,multi-band} & $25.3\pm2.6\%$ & 109 & $3.25\pm0.31$ & $89.9\pm8.6$ & $88.1\pm8.4$ \\
\enddata
\end{deluxetable*}

\begin{deluxetable*}{c r r r r r}
\tablewidth{0pt}
\tablecaption{Detection metrics for simulated SGRB afterglows in Argus.\label{tab:metrics_results_argus_sgrb}}
\tablehead{
\colhead{Metric} &
\colhead{Efficiency} &
\colhead{$N_{\rm detected}$} &
\colhead{$\Re$ (deg$^{-2}$ yr$^{-1}$)\,$\times10^3$} &
\colhead{$\Re_{\rm detection}$ (yr$^{-1}$)} &
\colhead{$\Re_{\rm detection,GRB}$ (yr$^{-1}$)}
}
\startdata
Swift/BAT & -- & 781 & $1.35\pm0.05$ & -- & $(7.6\pm0.3)$\\
\hline
\texttt{1det} & $13.1\pm1.2\%$ & 152 & $0.26\pm0.02$ & $7.3\pm0.6$ & $1.0\pm0.1$ \\
\texttt{2det} & $12.6\pm1.2\%$ & 147 & $0.25\pm0.02$ & $7.1\pm0.6$ & $1.0\pm0.1$ \\
\texttt{2det,band} & $9.7\pm1.1\%$ & 113 & $0.20\pm0.02$ & $5.4\pm0.5$ & $0.7\pm0.1$ \\
\texttt{2det,multi-band} & $12.0\pm1.2\%$ & 140 & $0.24\pm0.02$ & $6.7\pm0.6$ & $0.9\pm0.1$ \\
\texttt{fade} & $7.7\pm1.0\%$ & 90 & $0.16\pm0.02$ & $4.3\pm0.5$ & $0.6\pm0.1$ \\
\texttt{fade,multi-band} & $7.7\pm1.0\%$ & 90 & $0.16\pm0.02$ & $4.3\pm0.5$ & $0.6\pm0.1$ \\
\hline
Fermi/GBM & -- & 865 & $1.50\pm0.05$ & -- & $39\pm1$\\
\hline
\texttt{1det} & $13.7\pm1.2\%$ & 176 & $0.31\pm0.02$ & $8.4\pm0.6$ & $5.4\pm0.4$ \\
\texttt{2det} & $13.2\pm1.2\%$ & 170 & $0.29\pm0.02$ & $8.2\pm0.6$ & $5.2\pm0.4$ \\
\texttt{2det,band} & $10.2\pm1.0\%$ & 131 & $0.23\pm0.02$ & $6.3\pm0.5$ & $4.0\pm0.4$ \\
\texttt{2det,multi-band} & $12.6\pm1.1\%$ & 163 & $0.28\pm0.02$ & $7.8\pm0.6$ & $5.0\pm0.4$ \\
\texttt{fade} & $8.1\pm0.9\%$ & 105 & $0.18\pm0.02$ & $5.0\pm0.5$ & $3.2\pm0.3$ \\
\texttt{fade,multi-band} & $8.1\pm0.9\%$ & 105 & $0.18\pm0.02$ & $5.0\pm0.5$ & $3.2\pm0.3$ \\
\hline
StarBurst & -- & 2490 & $4.32\pm0.09$ & -- & $96\pm2$\\
\hline
\texttt{1det} & $9.4\pm0.6\%$ & 350 & $0.61\pm0.03$ & $16.8\pm0.9$ & $9.1\pm0.5$ \\
\texttt{2det} & $9.0\pm0.6\%$ & 333 & $0.58\pm0.03$ & $16.0\pm0.9$ & $8.6\pm0.5$ \\
\texttt{2det,band} & $7.2\pm0.5\%$ & 268 & $0.46\pm0.03$ & $12.9\pm0.8$ & $7.0\pm0.4$ \\
\texttt{2det,multi-band} & $8.6\pm0.6\%$ & 318 & $0.55\pm0.03$ & $15.3\pm0.9$ & $8.3\pm0.5$ \\
\texttt{fade} & $5.6\pm0.5\%$ & 209 & $0.36\pm0.03$ & $10.0\pm0.7$ & $5.4\pm0.4$ \\
\texttt{fade,multi-band} & $5.6\pm0.5\%$ & 208 & $0.36\pm0.03$ & $10.0\pm0.7$ & $5.4\pm0.4$ \\
\hline
MoonBEAM & -- & 1017 & $1.76\pm0.06$ & -- & $71\pm2$\\
\hline
\texttt{1det} & $12.5\pm1.0\%$ & 189 & $0.33\pm0.02$ & $9.1\pm0.7$ & $8.9\pm0.6$ \\
\texttt{2det} & $12.0\pm1.0\%$ & 182 & $0.32\pm0.02$ & $8.7\pm0.6$ & $8.6\pm0.6$ \\
\texttt{2det,band} & $9.1\pm0.9\%$ & 138 & $0.24\pm0.02$ & $6.6\pm0.6$ & $6.5\pm0.6$ \\
\texttt{2det,multi-band} & $11.5\pm1.0\%$ & 174 & $0.30\pm0.02$ & $8.3\pm0.6$ & $8.2\pm0.6$ \\
\texttt{fade} & $7.1\pm0.8\%$ & 108 & $0.19\pm0.02$ & $5.2\pm0.5$ & $5.1\pm0.5$ \\
\texttt{fade,multi-band} & $7.1\pm0.8\%$ & 108 & $0.19\pm0.02$ & $5.2\pm0.5$ & $5.1\pm0.5$ \\
\enddata
\end{deluxetable*}

\begin{deluxetable*}{c r r r r r}

\tablewidth{0pt}
\tablecaption{Detection metrics for simulated LGRB afterglows in DSA.\label{tab:metrics_results_DSA}}
\tablehead{
\colhead{Metric} &
\colhead{Efficiency} &
\colhead{$N_{\rm detected}$} &
\colhead{$\Re$ (deg$^{-2}$ yr$^{-1}$)\,$\times10^3$} &
\colhead{$\Re_{\rm detection}$ (yr$^{-1}$)} &
\colhead{$\Re_{\rm detection,GRB}$ (yr$^{-1}$)}
}
\startdata
Swift/BAT & -- & 423 & $12.60\pm0.61$ & -- & $71\pm3$\\
\hline
\texttt{1det} & $53.8\pm2.4\%$ & 345 & $10.27\pm0.61$ & $279.7\pm16.7$ & $38.2\pm2.3$ \\
\texttt{2det} & $45.3\pm2.4\%$ & 290 & $8.65\pm0.61$ & $235.5\pm16.7$ & $32.2\pm2.3$ \\
\texttt{3det} & $38.3\pm2.4\%$ & 246 & $7.31\pm0.61$ & $199.1\pm16.7$ & $27.2\pm2.3$ \\
\hline
Fermi/GBM & -- & 250 & $7.45\pm0.47$ & -- & $196\pm12$\\
\hline
\texttt{1det} & $57.9\pm3.1\%$ & 219 & $6.53\pm0.47$ & $177.7\pm12.8$ & $113.3\pm8.2$ \\
\texttt{2det} & $50.7\pm3.2\%$ & 192 & $5.72\pm0.47$ & $155.6\pm12.8$ & $99.2\pm8.2$ \\
\texttt{3det} & $43.8\pm3.1\%$ & 166 & $4.94\pm0.47$ & $134.5\pm12.8$ & $85.8\pm8.2$ \\
\hline
StarBurst & -- & 608 & $18.11\pm0.73$ & -- & $404\pm16$\\
\hline
\texttt{1det} & $54.5\pm2.0\%$ & 502 & $14.96\pm0.73$ & $407.2\pm20.0$ & $220.3\pm10.8$ \\
\texttt{2det} & $46.6\pm2.0\%$ & 430 & $12.79\pm0.73$ & $348.4\pm20.0$ & $188.5\pm10.8$ \\
\texttt{3det} & $39.2\pm2.0\%$ & 361 & $10.75\pm0.73$ & $292.6\pm20.0$ & $158.3\pm10.8$ \\
\hline
MoonBEAM & -- & 289 & $8.61\pm0.51$ & -- & $348\pm20$\\
\hline
\texttt{1det} & $59.1\pm2.9\%$ & 259 & $7.71\pm0.51$ & $210.0\pm13.8$ & $205.8\pm13.5$ \\
\texttt{2det} & $52.7\pm2.9\%$ & 231 & $6.88\pm0.51$ & $187.2\pm13.8$ & $183.5\pm13.5$ \\
\texttt{3det} & $46.1\pm2.9\%$ & 202 & $6.01\pm0.51$ & $163.7\pm13.8$ & $160.4\pm13.5$ \\
\enddata
\end{deluxetable*}

\begin{deluxetable*}{c r r r r r}

\tablewidth{0pt}
\tablecaption{Detection metrics for simulated SGRB afterglows in DSA.\label{tab:metrics_results_DSA_sgrb}}
\tablehead{
\colhead{Metric} &
\colhead{Efficiency} &
\colhead{$N_{\rm detected}$} &
\colhead{$\Re$ (deg$^{-2}$ yr$^{-1}$)\,$\times10^3$} &
\colhead{$\Re_{\rm detection}$ (yr$^{-1}$)} &
\colhead{$\Re_{\rm detection,GRB}$ (yr$^{-1}$)}
}
\startdata
Swift/BAT & -- & 781 & $1.35\pm0.05$ & -- & $7.6\pm0.3$\\
\hline
\texttt{1det} & $37.4\pm1.7\%$ & 443 & $0.77\pm0.05$ & $20.9\pm1.3$ & $2.9\pm0.2$ \\
\texttt{2det} & $13.4\pm1.2\%$ & 158 & $0.27\pm0.05$ & $7.5\pm1.3$ & $1.0\pm0.2$ \\
\texttt{3det} & $5.5\pm0.8\%$ & 65 & $0.11\pm0.05$ & $3.1\pm1.3$ & $0.4\pm0.2$ \\
\hline
Fermi/GBM & -- & 865 & $1.50\pm0.05$ & -- & $39\pm1$\\
\hline
\texttt{1det} & $40.2\pm1.7\%$ & 527 & $0.91\pm0.05$ & $24.9\pm1.4$ & $15.9\pm0.9$ \\
\texttt{2det} & $14.7\pm1.2\%$ & 193 & $0.33\pm0.05$ & $9.1\pm1.4$ & $5.8\pm0.9$ \\
\texttt{3det} & $5.9\pm0.8\%$ & 78 & $0.13\pm0.05$ & $3.7\pm1.4$ & $2.3\pm0.9$ \\
\hline
StarBurst & -- & 2490 & $4.32\pm0.09$ & -- & $96\pm2$\\
\hline
\texttt{1det} & $28.4\pm0.9\%$ & 1072 & $1.86\pm0.09$ & $50.6\pm2.4$ & $27.4\pm1.3$ \\
\texttt{2det} & $9.0\pm0.6\%$ & 341 & $0.59\pm0.09$ & $16.1\pm2.4$ & $8.7\pm1.3$ \\
\texttt{3det} & $3.2\pm0.4\%$ & 119 & $0.21\pm0.09$ & $5.6\pm2.4$ & $3.0\pm1.3$ \\
\hline
MoonBEAM & -- & 1017 & $1.76\pm0.06$ & -- & $71\pm2$\\
\hline
\texttt{1det} & $37.9\pm1.5\%$ & 584 & $1.01\pm0.06$ & $27.6\pm1.5$ & $27.0\pm1.5$ \\
\texttt{2det} & $13.5\pm1.1\%$ & 208 & $0.36\pm0.06$ & $9.8\pm1.5$ & $9.6\pm1.5$ \\
\texttt{3det} & $5.4\pm0.7\%$ & 84 & $0.14\pm0.06$ & $3.9\pm1.5$ & $3.9\pm1.5$ \\
\enddata
\end{deluxetable*}

Rates of afterglows associated with GRB detections with future missions, MoonBEAM and StarBurst, present a new paradigm in GRB science with joint detections between GRB monitors and synoptic surveys occurring over once a week with Argus and over twice a week with DSA.  Even without these new facilities, the rate of afterglow discovery with DSA and Argus presents a new paradigm where these surveys will provide afterglow discoveries, without requiring any dedicated resources, as or more frequently than can currently be achieved with a triggered approach.   Because of the transformational discovery abilities of Argus, new space-based GRB monitors may no longer need to be designed to prioritize rapid reporting of precise localizations.  Conversely, due to the comparatively low cadence of the DSA all-sky survey, radio detections will not be identified for months--years (see Fig.~\ref{fig:cumulative_min_time}.  Additionally, it may be difficult to distinguish their nature from contaminants such as AGN without precise localizations.

\begin{figure*}
    \centering
    \includegraphics[width=\linewidth,trim=5 8 5 5, clip]{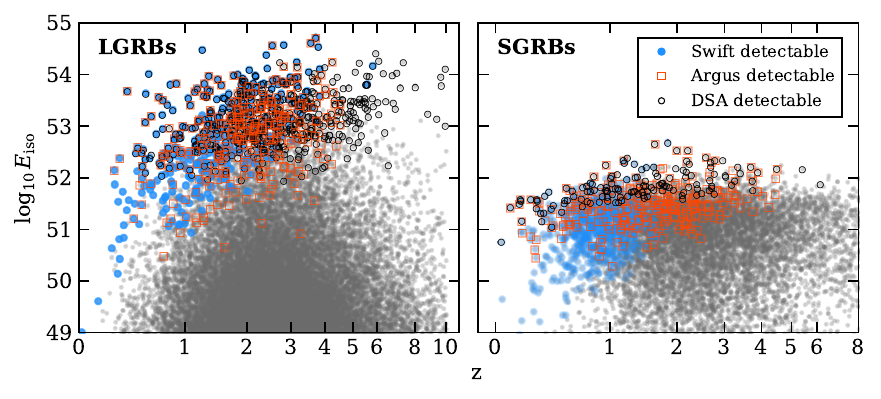}
    \caption{Isotropic equivalent energy release versus redshift for our population of synthetic GRBs.  The left panel shows the LGRB population and the right panel shows the SGRB population.  The grey points indicate all simulated GRBs, the blue circles indicate GRBs detected with Swift/BAT, the orange squares indicate GRBs with afterglows detected with Argus and the black rings indicate GRBs with afterglows detected with DSA.}
    \label{fig:Eiso_vs_z}
\end{figure*}

Without requiring GRB counterparts, our results show that DSA and Argus will likely yield order $>10^2$ LGRB afterglows per year, exceeding the current rate with global follow-up.  While these will be more difficult to identify from survey data, they present a new regime where afterglows will be routinely identified without high-energy triggers.  Fig~\ref{fig:Eiso_vs_z} shows the $\Eiso$--$z$ distribution for DSA and Argus detected afterglows in addition to Swift/BAT.  We find that DSA is limited to only the high $\Eiso$ bursts and that Argus and DSA can detect afterglows at higher redshifts than Swift/BAT.  This is likely due to $E_p$ being redshifted out of Swift/BAT's energy band.  This suggests a promising future for the detection of a population of ``orphan'' afterglows (i.e. those intrinsically without bright GRB emission) with both DSA and Argus, although we caution that the GS22 population is fit to observed GRB properties and may be unrealistic at low $\Eiso$ and $\Gamma_0$ values, where a GRB would be undetectable.  We therefore leave a detailed discussion of this avenue to future work.

From Tables~\ref{tab:metrics_results_argus_sgrb} and \ref{tab:metrics_results_DSA_sgrb} we find that the SGRB afterglow detection rate with DSA and Argus is 5--10\% that of the LGRB afterglow detection rate.  We expect that the rate of SGRB afterglows, independent of GRB triggers, is $10\pm1$ per year with Argus and $6\pm3$ per year with DSA.  However, a GRB association would be usually required to identify afterglows as likely originating from a neutron star merger.  This results in a rate of 2--3 events per year with both Argus and DSA with accompanying Fermi/GBM-detected GRBs.  These results are comparable to the current rate of SGRB afterglow discovery via global follow-up \citep{2015ApJ...815..102F}.

\subsection{Light curve sampling}\label{sec:lc_sampling}

\begin{figure}
    \centering
    \includegraphics[width=\linewidth]{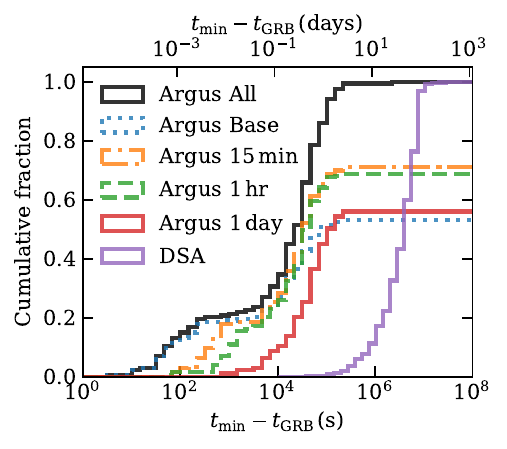}
    \caption{Cumulative distributions of the time between an LGRB detection and the first detection of its afterglow. Distributions are shown for each Argus cadence analyzed in this work; baseline (1\,second and 1\,minute), 15\,min, 1\,hr and 1\,day in addition to DSA and are expressed as a fraction of the total number of simulated detections.}
    \label{fig:cumulative_min_time}
\end{figure}

In Fig.~\ref{fig:lightcurve_example}, we display an example light curve for a bright LGRB afterglow with prominent reverse shock emission which is detected in our simulations by the Argus Array. The early optical flash and rise of the forward-shock emission will be routinely sampled -- \fracPrepeak~of LGRB events detected in our Argus sample have at least one pre-peak detection.  Fractions of the duration of the interval between the LGRB and the time of the first Argus detection is shown in Fig.~\ref{fig:cumulative_min_time}.  In the Argus simulations, \fracDayNotBase~of afterglows are detectable in the day-cadence data but not the baseline cadence, \fracBaseNotDay~are detectable in the baseline cadence but not the day cadence and \fracBaseAndDay~are detectable in both.  We find that the 15\,min cadence accounts for \fracFifteen~of Argus-detected afterglows, yielding the highest detection rates.

\begin{figure*}
    \centering
    \includegraphics[width=\linewidth,trim=30 60 30 60, clip]{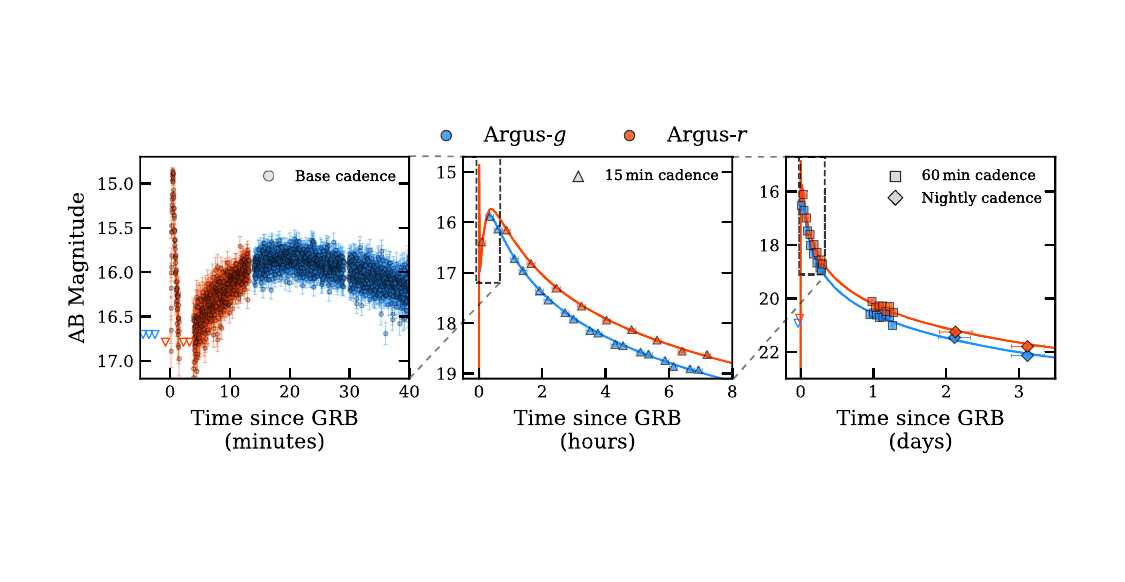}
    \caption{An example of a LGRB afterglow model with a prominent, early reverse shock, showing simulated observations at the base cadence (left panel), 15-minute coadds (center panel), and 60-minute and nightly coadds (right panel).  The injected light curve is shown with solid lines and its parameters are: $z = 1.47$, $E_{\mathrm{iso}} = 7.98 \times 10^{53}$\,erg, $n_0 = 0.64$\,cm$^{-3}$, $\theta_c = 12.42^\circ$, $p = 2.58$, $\log \epsilon_e = -1.02$, $\log \epsilon_B = -4.83$, $t_{\mathrm{j}} = 3.24$\,s, $\log R_b = 3.85$ and $A_V = 0.24$.}
    \label{fig:example_argus_lightcurve}
\end{figure*}

The fraction of LGRB afterglows in our Argus sample with reverse shock emission peaking brighter than the forward shock peaks is $\sim3$\%.  However, for afterglows peaking brighter than $R_c=16$\,AB mag the fraction is $\sim10$\% which is consistent with the results of \citet{2014ApJ...785...84J}.  This is a consequence of the scaling $f_{\rm peak,rvs}/f_{\rm peak,fwd}\propto\Gamma_0R_B^2$ \citep{2003ApJ...582L..75K,2003ApJ...595..950Z,2020ApJ...895...94Y} and the observed correlation (which was imposed on our synthetic population) $E_{\rm iso}\propto\Gamma_0^\alpha$ \citep{2010ApJ...725.2209L,2012ApJ...751...49L,2012MNRAS.420..483G,2018A&A...609A.112G}.  Therefore, under this correlation, more energetic LGRBs and with more luminous forward shock emission have relatively brighter reverse shock emission.  Conversely, some works dispute the existence of an intrinsic $E_{\rm iso}$--$\Gamma_0$ correlation on the basis of Malmquist bias \citep[e.g.][]{2012ApJ...747..146K,2015MNRAS.449L...6C}. Nonetheless, Argus will disentangle many of these biases by sampling the rise phase of the afterglow light curves.  It will do this by; measuring the fraction of detectable reverse shock components and its correlation with $E_{\rm iso}$, directly measuring $\Gamma_0$ for a statistical sample of afterglows both with and without high-energy triggers and determining whether optical flashes are caused by the low-energy tail of the prompt emission using second and minute-cadence photometry \citep[e.g.][]{1999MNRAS.306L..39M,2005Natur.435..178V,2010ApJ...719L..10B,2026arXiv260305608J}. 

At radio wavelengths, the DSA observations provide an excellent avenue to sample the late-time afterglow emission through late times ($\sim$yrs). Fig.~\ref{fig:lightcurve_example} hints at the information that could be extracted from afterglows jointly detected with DSA and Argus. Afterglows may be difficult to identify independently in the DSA alert stream due to the lack of cadence and source confusion.  In Fig~\ref{fig:cumulative_min_time}, we show that most afterglows detected with DSA will have their first detection $>50$\,days post-burst. The key benefit comes from treating Argus and DSA as a single observatory system and producing a single multi-wavelength dataset that spans eight orders of magnitude in both timescale and frequency for simultaneous detections. By fitting afterglow models to the multi-wavelength light curve, we will be able to break afterglow model degeneracies that arise from only considering a single wavelength regime. This approach is not particularly novel and has been applied to a variety of GRBs in the past \citep[e.g.][]{2001ApJ...554..667P,2003ApJ...597..459Y}, but the key breakthrough is that conducting it at population scales will no longer be limited by follow-up resources.

\subsection{Comparisons with other time-domain surveys}

Assuming the \texttt{fade} metric, for events that can be reliably identified from the Argus alert stream,  the serendipitous detection rate of LGRB afterglows in Argus of  $\fermiargusrate$\,yr$^{-1}$ for Fermi/GBM and $\swiftargusrate$\,yr$^{-1}$ for Swift/BAT eclipses that of ZTF with $\sim 1$--$2$\,yr$^{-1}$ \citep{2021ApJ...918...63A}.  Without coadded data, Argus' detection rate drops by only $\sim$46\%.  This is owed to its high cadence and sky-coverage compared to ZTF.

We follow the same procedure as in \S\ref{sec:argus_sims}, with simulated observations from the NSF-DOE Vera C. Rubin observatory's Legacy Survey of Space and Time \citep[LSST;][]{2019ApJ...873..111I}, with version \texttt{baseline\_v5.1.1\_10yrs} \citep{2021ApJS..253...31L}.  For LGRBs detected by Fermi/GBM, we find that the Argus Array will yield a detection rate, \lsstFactorOneDet, \lsstFactorTwoDetSame~and \lsstFactorFade~times larger for the \texttt{1det}, \texttt{2det,band} and \texttt{fade} metrics respectively.  While Rubin reaches deeper than Argus, in more filters,  Argus outperforms Rubin in detecting afterglows due to its high cadence. Afterglows' duration above half-maximum light of $t_{1/2}<1$\,day \citep{2022ApJ...938...85H} dwarfs the average return visit time for $r$-band in \texttt{baseline\_v5.1.1\_10yrs} of 2--4 days. In contrast, Rubin and Argus will typically have $\sim$580 square degrees of overlap in their footprint on a given night.  Rubin will perform deeper, multi-band imaging into UV and NIR wavelengths which will better enable discoveries of supernova and kilonova emission components at late times.  This capability will be analyzed in detail in a future work, A. Tartaglia et al., in preparation (2026).

In radio wavelengths, currently there are three wide-field surveys capable of discovering GRB afterglows serendipitously. The Rapid ASKAP Continuum Survey \citep{2020PASA...37...48M} has already yielded one GRB detection \citep{2021MNRAS.503.1847L}, while the VLA Sky Survey has discovered several synchrotron afterglows \citep{2025ApJ...995...61S,2025PASP..137h4102S}, and one likely GRB afterglow \citep{2021ApJ...923L..24S,2021ApJ...907...60M}. Both surveys are limited by their relatively long observing cadence, meaning that afterglows would typically manifest as single detections, indistinguishable from shorter-lived transients without comprehensive follow-up observations. The VAST extragalactic survey has a $\sim$2 month observing cadence, which is more suited to the expected timescales of GRB afterglows, albeit covering a smaller fraction of the sky compared to RACS/VLASS. While there are no reports of VAST-detected GRBs yet, we have carried out a similar analysis to the one presented in this work for DSA, and find that $\vastafterglows\,\rm yr^{-1}$ afterglows should be present in VAST. However, given the sensitivity (5-$\sigma$ threshold $1.25\,$mJy), a lot of these are detections close to the noise floor. For example, $<$20\% of the VAST-detected GRBs in our simulations have peak-detected flux density of $>$5\,mJy (SNR of 20), to independently identify an afterglow candidate at high confidence. This means that, although detectable, independent VAST discoveries have to contend with distinguishing between GRB afterglows and other sources of confusion, like AGN variability, which will be difficult, and hence only some subset of afterglows present in the VAST survey may be reliably recovered.

\begin{figure}
    \centering
    \includegraphics[width=\linewidth,trim=5 5 5 5, clip]{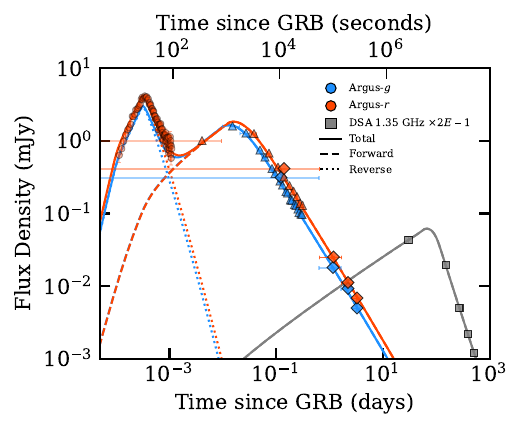}
    \caption{The same model as in Fig.~\ref{fig:example_argus_lightcurve} with simulated DSA observations, shown as grey squares.  We plot the injected light curve including the reverse and forward shock emission components.  Simulated Argus observations with the baseline (second) cadence data are shown with circles, 15-minute cadence data are shown as triangles and day-cadence data are shown with diamonds.}
    \label{fig:lightcurve_example}
\end{figure}

{
\subsection{Prospects for light curve filtering}

Facilities like Swift and SVOM, providing arcminute-scale localizations for GRBs, makes identifying their multi-wavelength counterparts trivial and can usually be achieved with relatively narrow field telescopes.  Argus and DSA's strength lies in serendipitously detecting afterglows without the need for well-localized GRB detections.  The challenge for this approach is filtering out false positives.  Taking all transients first detected after the GRB detection and only those in the localization region will be likely to yield many candidates. More advanced filtering will therefore be required to positively identify counterparts.  

Depending on the cadence, filtering Argus alerts will likely take on different forms.  For promptly identifying GRB counterparts within seconds--minutes, candidates will be dominated by stellar flares \citep[e.g.][]{2020MNRAS.491.5852A,2024MNRAS.531.4836F} and glints from satellite debris \citep{2020ApJ...903L..27C,2021MNRAS.505.2477N,2025ApJ...994..175T,2025MNRAS.543.3915G}.  The fade rate metric in this case will not be particularly effective as, at early times, afterglows can have multi-peaked behavior owing to prompt and reverse shock emission \citep[e.g.][]{1999ApJ...517L.109S,2009ApJ...691..723B,2011A&A...528A..15G,2013ApJ...774..114J,2014Sci...343...38V,2017Natur.547..425T,2019ApJ...879L..26F,2023NatAs...7..843O,1999MNRAS.306L..39M,2005Natur.435..178V,2008Natur.455..183R,2010ApJ...719L..10B,2026arXiv260305608J} and contaminants will be also be fast fading.  To mitigate this, cross-matching with deep photometric surveys such as LSST \citep{2019ApJ...873..111I}, Gaia \citep{2023A&A...674A...1G} and the DESI Legacy Surveys \citep{2019AJ....157..168D} will be crucial for removing stellar contaminants.  Requiring a red color ($g-r>0$) would also be effective in rejecting stellar flares, although this would introduce latencies of up to 15\,minutes to detect the source in another filter.  To remove satellite debris, cutting candidates without multiple detections, non-PSF detections and alerts linked to tracklets may be effective.  Adding additional cuts necessarily increases latency between the GRB detection and the identification of the counterpart.  This motivates a tiered system where increasingly stringent cuts are introduced.  Candidates can then be disseminated rapidly after the GRB detection but high purity can be established with time.

Identifying afterglow candidates on longer timescales will be important, for example, where a GRB was detected without simultaneous coverage from Argus or for untargeted searches for orphan afterglows.  Here cataclysmic variables (CVs) will dominate contaminants.  Here, cross-matching to rule out candidates coincident with stars and flag those coincident with galaxies in addition to cutting on both fast fade rates and Galactic extinction corrected red color is likely to be effective for removing contaminants \citep[e.g.][]{2021ApJ...918...63A}.  From the analysis in \S\ref{sec:lc_sampling}, we conclude that applying these metrics to all of the coadd intervals simultaneously will maximize recovery efficiency, favoring the 15-minute coadds.

The approach for identifying afterglow candidates with DSA will be largely unchanged from previous efforts \citep[e.g.][]{2020PASP..132c5001L,2023MNRAS.523.4029L} albeit with a larger data volume.  We therefore predict that radio variability from Active Galactic Nuclei (AGN) will be the primary source of contaminants.  The relatively low cadence of the DSA all-sky survey will present challenges in distinguishing AGN from afterglows with many months going by before an afterglow can be distinguished from AGN variability from its light curve alone.  While cross-matching with known AGN will be effective, the unprecedented depth of DSA will present a large population of uncatalogued AGN.  Therefore, an AGN catalog generated from DSA data itself will be invaluable for filtering.
}

\subsection{Implications for future follow-up campaigns}

Our results show that Argus and DSA will generate afterglow discoveries, in their respective wavelengths, through their regular survey operations at rate that rivals what is currently possible with a triggered approach, due largely to the fact that the majority of detected GRBs are poorly localized (e.g. Fermi).  Argus will sample their early light curves, at a combination of depth and cadence that is impossible for most other optical facilities.  In addition, after Swift concludes science operations, Argus may be relied on for localizing and sampling early GRB afterglow light curves. However, Argus' footprint is limited to the Northern hemisphere and by the weather and light conditions at the Argus site.

Similarly, at radio wavelengths, our results show that DSA will amass a sample of radio afterglows larger than all radio-discovered afterglows to date. In addition to its ability to uniquely sample the late-time afterglow emission in a large sample of GRBs, it opens up a new parameter space to independently discover afterglows that are fainter at other wavelengths. DSA's single epoch sensitivity, which rivals the capabilities of most current telescopes, will complement targeted efforts in large samples. However, this might be limited to sampling the late-time afterglow emission. Given its 4-month cadence, targeted observations might still be required in individual sources that evolve on faster timescales. 

\begin{figure}
    \centering
    \includegraphics[width=\linewidth,trim=7 7 7 10, clip]{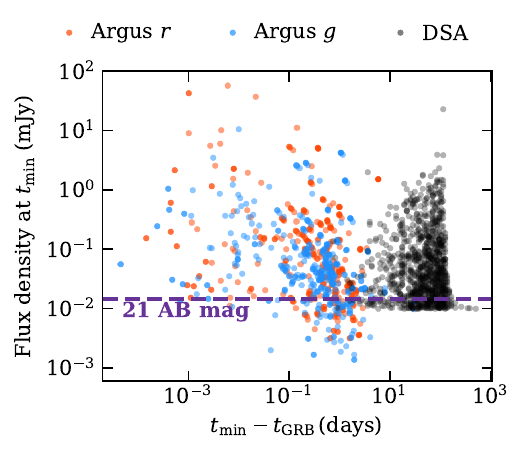}
    \caption{Time and flux density of afterglow first 5-$\sigma$ detections in Argus $g$ and $r$-bands in addition to DSA.  The horizontal, dotted line indicates 21\,AB mag, above which spectroscopy is feasible.}
    \label{fig:flux_vs_min_time}
\end{figure}

Further observations from other facilities will be required to fully exploit the multi-wavelength scientific potential of a given event.  Specifically, Argus conducts observations in only optical band passes, $g$ and $r$.  We therefore highlight the utility of follow-up of Argus-detected afterglows in the form of infrared and ultraviolet imaging to constrain thermal emission and rapid-response spectroscopy to measure absorption features. In Fig~\ref{fig:flux_vs_min_time}, we illustrate that spectroscopy will be feasibly for the majority of Argus-detected afterglows at the time of the first detection.  DSA, while sensitive, lacks the cadence to independently establish the afterglow nature on a timescale over which higher energy emission evolves. Additionally, in general, multiple epochs of radio data might be needed in order to establish the afterglow nature and confidently rule out sources of confusion like AGN variability. Our results in Fig.~\ref{fig:flux_vs_min_time} show that these follow-up radio observations will be feasible for most afterglows after an initial detection with DSA. However, there can be extragalactic sources showing extreme variability (a factor of $>5$ between epochs), which can immediately hint at afterglow emission. High cadence radio and multi-wavelength observations, either on Argus or DSA discoveries, will therefore be crucial for constraining jet and environment properties. 

\section{Conclusion}\label{sec:conclusion}

With the Argus Array and the Deep Synoptic array, the Eric and Wendy Schmidt Observatory System will dramatically increase the rates of serendipitously discovered afterglows, rivaling what can be achieved currently with triggered follow-up of GRBs.

We have simulated DSA and Argus observations of a synthetic population of GRB afterglows with parameters matching those of the observed population of X-ray, optical and radio light curves. We  compute rates of afterglow discovery with DSA and Argus for GRBs detected with a range of instruments. We find that $\fermiarguseff$ and $\fermidsaeff$ of GRBs detected with Fermi/GBM will have counterparts in Argus and DSA respectively.

We also include both planned and proposed missions, such as the Moon Burst Energetics All-sky Monitor (MoonBEAM) mission concept, and the StarBurst Multimessenger Pioneer (to launch in 2028). GRBs detected with MoonBEAM ($\moonbeamargusrate$ with Argus and $\moonbeamdsarate$ with DSA per year) and StarBurst ($\starburstargusrate$ with Argus and $\starburstdsarate$ with DSA per year) will yield greater than weekly afterglow discovery in Argus and almost twice weekly in DSA.  We consider these rates to be lower limits as they do not include events without GRB detections including low-Lorentz factor and off-axis events.

As a result, Argus will supplement efforts to localize GRBs, currently done with Swift's Ultraviolet-Optical and X-ray telescopes in addition to ground-based follow-up.  Without precise localization, identifying afterglows in the Argus data stream will be hindered by contaminants such as stellar flares and cataclysmic variables.  However, with appropriate cross-matching and filtering, we predict afterglow candidates will often be distinguishable from contaminants within an hour of discovery. Similarly, DSA will supplement radio afterglow discovery efforts although its cadence and northern footprint will still require triggered follow-up to discover them across the sky in a timely manner.  Without precise localization, distinguishing radio afterglows from AGN variability will continue to be a challenge in the DSA data stream.

Of the afterglows detected with Argus, $\sim\fracPrepeak$~will have pre-peak detections.  This will enable measurements of initial Lorentz factors of GRBs at a population-level.  It will also routinely yield detections of optical flashes, preceding the forward shock emission.  This will help disentangle the reverse shock and the low energy tail of the prompt emission as progenitors for these signatures. Joint-observations of afterglows with DSA and Argus will yield light curves with temporal sampling spanning $\sim$8 orders of magnitude, providing rich datasets to study jet geometry and progenitor environments.

Triggered follow-up of afterglows detected with Argus and DSA in the form of rapid response spectroscopy, ultraviolet and infrared photometry and high-cadence and multi-frequency radio observations will be crucial.  We therefore highlight the importance of developing methods for identifying them promptly in Argus and DSA survey data.

\begin{acknowledgments}

I.A. is supported by the National Science Foundation award AST 2505775, NASA grant 24-ADAP24-0159, Scialog award SA-LSST-2024-102a and LSST-2025-112b. B.O. is supported by the McWilliams Postdoctoral Fellowship in the McWilliams Center for Cosmology and Astrophysics at Carnegie Mellon University.  JC and HC are supported by Schmidt Sciences and Alex Gerko.

This work made use of data supplied by the UK \textit{Swift} Science Data Centre at the University of Leicester. 
This research has made use of the XRT Data Analysis Software (XRTDAS) developed under the responsibility of the ASI Science Data Center (ASDC), Italy. 
This research has made use of data and/or software provided by the High Energy Astrophysics Science Archive Research Center (HEASARC), which is a service of the Astrophysics Science Division at NASA/GSFC. This research has made use of the Astrophysics Data System, funded by NASA under Cooperative Agreement 80NSSC21M00561.


\end{acknowledgments}



\software{
\texttt{astropy} \citep{2013A&A...558A..33A,2018AJ....156..123A,2022ApJ...935..167A},
\texttt{dustmaps} \citep{2018JOSS....3..695G},
\texttt{numpy} \citep{2020Natur.585..357H},
\texttt{matplotlib} \citep{2007CSE.....9...90H},
\texttt{pandas} \citep{2022zndo...3509134T},
\texttt{redback} \citep{2024MNRAS.531.1203S},
\texttt{scipy} \citep{2020NatMe..17..261V},
\texttt{sncosmo} \citep{2016ascl.soft11017B},
\texttt{VegasAfterglow} \citep{2026JHEAp..5000490W}
}



\bibliography{ref}{}
\bibliographystyle{aasjournalv7}

\end{document}